\documentstyle[12pt]{article}
\newcommand{\ifm}[1]{\relax\ifmmode #1\else $#1$\hskip 0.15cm\fi}
\newcommand{\etal}{\mbox{\it et al.}}
\newcommand{\ie}{\mbox{\it i.e.}}

\newcommand{\gev}{\ifm{\mbox{GeV}}}

\newcommand{\tbar}{\ifm{\overline{t}}}

\newcommand{\ttbar}{\ifm{t\tbar}}

\newcommand{\qqbar}{\ifm{q\qbar}}
\newcommand{\p}[1]{\ifm{p_{_#1}}}
\newcommand{\px}{\p{x}}
\newcommand{\py}{\p{y}}

\newcommand{\pt}{\p{T}}

\newcommand{\mpt}{\mbox{$\rlap{\kern0.1em/}\pt$}}
\newcommand{\mpx}{\mbox{$\rlap{\kern0.1em/}\px$}}
\newcommand{\mpy}{\mbox{$\rlap{\kern0.1em/}\py$}}
\newcommand{\E}[1]{\ifm{E_#1}}
\newcommand{\Ex}{\E{x}}
\newcommand{\Ey}{\E{y}}

\newcommand{\Et}{\E{T}}
\newcommand{\ET}{\Et}
\newcommand{\mET}{\mbox{$\rlap{\kern0.1em/}\ET$}}
\newcommand{\mEx}{\mbox{$\rlap{\kern0.1em/}\Ex$}}
\newcommand{\mEy}{\mbox{$\rlap{\kern0.1em/}\Ey$}}

\newcommand{\be}{\begin{equation}}
\newcommand{\ee}{\end{equation}}
\newcommand{\bdm}{\begin{displaymath}}
\newcommand{\edm}{\end{displaymath}}

\newcommand{\Mean}[1]{{\mbox{$\left\langle #1 \right\rangle$}}}
\newcommand{\fracerr}[1]{\ifmmode
                    \frac{\delta #1}{#1}
               \else
                    \mbox{${\delta #1}/{#1}$}
               \fi}

\newcommand{\ttb}{\ifm{t\tbar}}

\newcommand{\cstt}{\ifm{\sigma(\ttb)}}
\newcommand{\csttp}{$\sigma(t \bar t)$}

\newcommand{\dzero}{\mbox{D\O}\hskip 0.14cm}

\newcommand{\tb}{\ifm{\tan\beta}}
\newcommand{\tbp}{$\tan\beta$}

\newcommand{\mHp}{$m_{H^{+}}$}
\newcommand{\mt}{\ifm{m_t}}
\newcommand{\mtp}{$m_{t}$}

\newcommand{\tHb}{\ifm{t \rightarrow H^+b}}
\newcommand{\tWb}{\ifm{t \rightarrow W^+b}}

\newcommand{\Missing}[2]{{\mbox{$#1\kern-0.57em\raise0.19ex\hbox{/}_{#2}$}}\ }

\newcommand{\vMissing}[2]{\ifmmode
            \vec{#1}\kern-0.57em\raise.19ex\hbox{/}_{#2}
         \else
            {{\mbox{$\vec{#1}\kern-0.57em\raise.19ex\hbox{/}_{#2}$}}\ }
         \fi}

\newcommand{\met}{{\it E \hspace{-4.2mm}/}\hspace{0.6mm}$_T$}

\def\pbarp{$\overline{p}p $}            
\def\qqbar{$q\overline{q}$}             
\def\D0{D\O}                            
\def\ipb{pb$^{-1}$}                     
\def\pt{$p_T$}                          
\def\etal{{\sl et al.}}                 
\def\met{\mbox{${\hbox{$E$\kern-0.6em\lower-.1ex\hbox{/}}}_T$ }} 
\def\metx{\mbox{${\hbox{$E$\kern-0.6em\lower-.1ex\hbox{/}}}_{x}$ }} 
\def\mety{\mbox{${\hbox{$E$\kern-0.6em\lower-.1ex\hbox{/}}}_{y}$ }} 
\def\d0draft{}

\def\mathunit#1{\mathop{\hbox{#1}}\mathclose{}\mathord{}}

\def\gev{\mathunit{GeV}}
\def\mev{\mathunit{MeV}}

\def\gevc{\gev \kern -1.7pt/ \kern -1.7pt c}

\def\mevc{\mev \kern -1.7pt/ \kern -1.7pt c}

\def\({\left(}
\def\){\right)}

\newcommand{\PRL}{{\em Phys. Rev. Lett.} }

\def\err#1#2#3 {{\it Erratum} {\bf#1},{\ #2} (19#3)}
\def\ib#1#2#3 {{\it ibid.} {\bf#1},{\ #2} (19#3)}
\def\nc#1#2#3 {Nuovo Cim. {\bf#1} ,#2(19#3)}
\def\nim#1#2#3 {Nucl. Instr. Meth. {\bf#1},{\ #2} (19#3)}
\def\np#1#2#3 {Nucl. Phys. {\bf#1},{\ #2} (19#3)}
\def\pl#1#2#3 {Phys. Lett. {\bf#1},{\ #2} (19#3)}
\def\prev#1#2#3 {Phys. Rev. {\bf#1},{\ #2} (19#3)}
\def\prl#1#2#3 {Phys. Rev. Lett. {\bf#1},{\ #2} (19#3)}
\def\rmp#1#2#3 {Rev. Mod. Phys. {\bf#1},{\ #2} (19#3)}
\def\zp#1#2#3 {Zeit. Phys. {\bf#1},{\ #2} (19#3)}
\input psfig

\def\Journal#1#2#3#4{{#1} {\bf #2}, #3 (#4)}

\def\NPB{Nucl. Phys. B}
\def\PRL{Phys. Rev. Lett.}

\begin{document}
\topskip 2cm
\begin{titlepage}

\begin{center}
{\large\bf D\O\ Findings on the Top Quark~\footnote{presented at the 1998
Rencontres de Physique de la Vall\'{e}e d'Aoste on Results and Perspectives
in Particle Physics, La Thuile, Italy, March 1--7, 1998.}} \\
\vspace{2.5cm}
{\large Boaz Klima} \\
\vspace{.5cm}
{\sl Fermilab }\\
\vspace{.5cm}
{\sl for the D\O\ collaboration }\\
\vspace{.5cm}
{\sl (e-mail: klima@fnal.gov) }\\
\vspace{2.5cm}
\vfil
\begin{abstract}

Recent results on top quark physics with the D\O\ experiment
in \pbarp \  collisions at $\sqrt{s} = 1.8$ \rm {TeV} for
an integrated luminosity of 125 \ipb are
reported.
The direct measurement of the top quark
mass uses single lepton and dilepton events, giving the result
$m_{top} = 172.1 \pm 7.1 \ {\rm GeV/c}^2$.
The measurement of the \ttbar  production cross section
includes analyses from 9 top decay channels: dilepton
(\ttbar $ \rightarrow e\mu$, $ee$, and $\mu\mu$), electron and neutrino
(\ttbar $ \rightarrow e\nu$), single
leptons (\ttbar $ \rightarrow e + {\rm jets}$,
\ttbar $ \rightarrow \mu + {\rm jets}$)
with and without $b$ tagging, and all-jets
(\ttbar $ \rightarrow 6 {\rm \hskip 2mm jets}$).
We measure the \ttbar  production cross section to be $5.9 \pm 1.7 \ \rm {pb}$
at $m_{top} = 172.1 \ {\rm GeV/c}^2$.
Combining the D\O\ and the CDF measurements of the top quark mass and
combining their \ttbar cross sections, in both cases taking into account
error correlations, yields unofficially
$\sigma_{t\overline{t}} = 6.7\pm1.3$ pb at
an averaged top quark mass of $m_{top} = 173.8 \pm 5.2$ GeV/c$^2$.
Preliminary results on a search for charged Higgs production in top events,
$\tHb$, are presented.

\end{abstract}

\end{center}
\end{titlepage}

\section{ Introduction }

Since the top quark was discovered~\cite{d0discov,cdfdiscov} by the D\O\ and CDF experiments
at the Fermilab Tevatron collider in 1995, much of the
effort has been focused
on using all available information on the cadidate events to measure
the \ttbar production cross section and
the top quark mass as precisely as possible.

In the Tevatron \pbarp \ collider
at a center of mass energy of $1.8 \ \rm {TeV}$ top quarks
are predominantly produced in pairs through
\qqbar \  annihilation ($\sim 90\%$)
or gluon fusion ($\sim 10\%$).
According to the standard
model a top quark decays almost 100\% of the time into a
$W$ and a $b$ quark. Each $W$ boson decays either into a charged lepton and a
neutrino or into a pair of quarks. Our analysis channels
are classified on the basis of the $W$ boson decay.
The cleanest channel
is the dilepton channel, \ttbar $ \rightarrow$ 2 leptons + 2 neutrinos + 2 jets,
where both $W$ bosons decay into leptons. The
branching ratio is $4\over81$ for \ttbar  decays into $e\mu$, $ee$,
and $\mu \mu$ channels.
The single lepton ($e$ or $\mu$) decay channel, where one $W$ boson decays
leptonically and the other $W$ boson decays hadronically, has a branching ratio of
$24\over81$, but it has a sizable background from
$W$+jets production. The all-jets decay channel, where both $W$ bosons decay
hadronically, has a branching ratio of
$36\over81$, but it has a huge QCD multijet background.

This paper reports the recent D\O\ results on the \ttbar production
cross section using the dilepton, single lepton~\cite{d0xsec},
and all-jets~\cite{d0alljets} decay channels,
and on the direct measurement
of the top quark mass using the single lepton~\cite{d0masslj,d0massPRD}
and dilepton events~\cite{d0massll}.

The results reported in this paper use the entire data sample
with an integrated luminosity of about 125  \ipb,
collected during the 1992-1996 collider run.
Since our report on the discovery of the top quark~\cite{d0discov}, our
data sample has more than doubled. D\O\ has optimized the analysis to maximize the
expected precision of the \ttbar production cross section measurement and
improved the techniques on the measurements of the top quark mass.

\section{ Measurement of \ttbar Production Cross Section}

The triggers and the particle identification algorithms used in these analyses
are described in detail
in D\O\ publications ~\cite{d0discov,searchprd}.

\subsection{ Dilepton Decay Channels}

The signature of \ttbar events in the dilepton decay channels is
two high $p_T$ isolated leptons, two or more jets and large missing $E_T$
(\met) due to the presence of the two neutrinos.

The offline
event selection requires that the isolated electrons have $E_T > 20 \ {\rm GeV}$
with $|\eta| < 2.5$ and that the isolated muons have $p_T > 15 \ {\rm GeV}$
with $|\eta| < 1.7$.
At least two jets are required to be reconstructed with a
transverse energy above $20 \ {\rm GeV}$ with $|\eta| < 2.5$.
All jets are reconstructed using a cone algorithm with radius
0.5 in $\eta$-$\phi$ space.
The \met is required to be above $25 \ {\rm GeV}$ for the $ee$ channel
and to be above $20 \ {\rm GeV}$ for the $e\mu$ channel.
In addition, we apply a cut on a variable $H_T$, which is
defined as the scalar sum of the $E_T$'s of all the jets in the event plus the
leading electron $E_T$.
We expect a higher $H_T$ for \ttbar events than for background
events. We require the $H_T$ to be above $120 \ {\rm GeV}$ for the
$ee$ and $e\mu$
channels, and to be above $100 \ {\rm GeV}$ for the $\mu\mu$ channel.

The background in these channels is mainly
from $Z^0$ decays, Drell-Yan, vector boson pair events and
events with misidentified leptons.

Five events passed the above selection criteria for the dilepton
decay channel, with estimated background $1.4 \pm 0.4$
events (see Table~\ref{tab:all}).

\subsection{ Single Lepton Decay Channels}

The signature of \ttbar  events in the single lepton channels is
a high $p_T$ isolated lepton, large \met and high jet multiplicity.
We divide the single lepton channel analysis into two
complementary and orthogonal analyses: an event shape analysis and a
$b$ tagging analysis.  The main sources of background in the
single lepton channels are $ W + \rm {jet} $ events with high jet
multiplicity  and multijet QCD events with fake leptons and \met due
to mismeasurement.

In the event shape analysis the \ttbar $ \rightarrow $ \rm ${e + jets} $
and \ttbar $ \rightarrow \mu + \rm { jets} $ events are selected by use of
topological and kinematic cuts.
The offline event selection
requires one isolated lepton
with $E_T > 20 \ {\rm GeV}$ with $|\eta_e| < 2.0$ or $|\eta_\mu| < 1.7$
and at least 4 jets with $E_T > 15 \ {\rm GeV}$ with $|\eta_{jet}| < 2.0$.
The \met requirement is
25 {\rm GeV} for the $e$ + jets channel and 20 {\rm GeV} for the $\mu$ + jets
channel.
The event shape analysis uses the global event variable aplanarity, $\cal A$,
which is defined as $3\over2$ of the smallest normalized eigenvalue of the
3-momentum tensor of the $W$ boson and jet momenta in the laboratory frame.
Background events typically have smaller ${\cal A}$
than signal events.
Another variable used in the event shape analysis is $H_T$, which is
defined as the sum of the $E_T$'s of all the jets in the event.
The \ttbar events are expected to have a higher $H_T$
than $W$+jets events.
In addition we also require the total leptonic $E_T$, $E_T^L$,
which is defined as
the sum of the charged lepton $E_T$ and \met,
to be above $60 \ {\rm GeV}$ to reject
the multijet QCD background. We require the pseudorapidity $\eta_W$ of the
$W$ boson that decays leptonically to have $|\eta_W| < 2.0$ to obtain
better agreement between background control samples from data
and the $W$+jets Monte Carlo (MC) samples~\cite{d0xsec}.

\vskip -0.5cm
\begin{figure}[htb]
\centerline{\hbox{
\psfig{figure=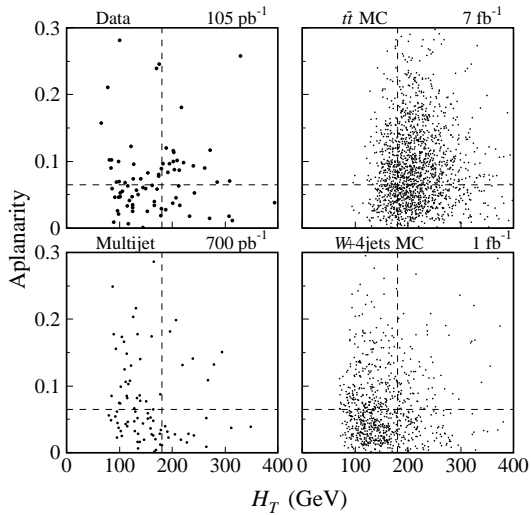,width=70.0mm}}}
\vskip -0.5cm
\caption{Aplanarity vs $H_T$ for single lepton data, \ttbar MC events,
multijet background and $W$+4 jets MC background.
The dashed lines indicate the cuts.}
\vskip -0.3cm
\label{fig:aht}
\end{figure}

Figure~\ref{fig:aht} shows scatter plots of $\cal A$ vs $H_T$ for
D\O\ single lepton data, \ttbar  ($m_{top} = 170 \ {\rm GeV/c}^2$)
MC events, and the two background sources: multijet background and
$W$ + 4 jets {\sc vecbos}~\cite{vecbos} MC events.
Based on our optimization procedure
using \ttbar MC events, we define the region
$ \cal {A} > $ 0.065 and $H_T > 180 \ {\rm GeV}$ as the signal region.

$W$ + jets background events and the multijet QCD events are the main source of
backgrounds in the lepton + jets channel. Since the number of $W$ + jets events
is expected to decrease exponentially as a function of the jet multiplicity
\cite{berends}, by fitting the number of $W$ + jets events at the lower
jet multiplicity and extrapolating it to high jet multiplicities,
we can estimate the number of $W$ + jets background events in the data sample.
We estimate the QCD background from the data itself using the measured jet
misidentification probability.

Nineteen events survived all the cuts in the event shape analysis for
the single lepton channel, with estimated background $8.7 \pm 1.7$
events (see Table~\ref{tab:all}).

In the $b$ tagging analysis,
the background for the single lepton channel can be significantly reduced
by requiring that one of the jets is tagged as a $b$-jet.
We tag $b$'s by detecting a muon in a jet.
About 20\% of \ttbar events have a
detectable $\mu$ in a jet compared to only about 2\% of the
$W+(\ge)$3jets background events.
A tag muon is
required to have $p_T^{\mu} > 4 \ {\rm GeV}$ and
the distance $\Delta R$  between the muon and a jet in the
      $\phi$-$\eta$ plane must be less than 0.5.

\begin{figure}[htb]
\centerline{\hbox{
\psfig{figure=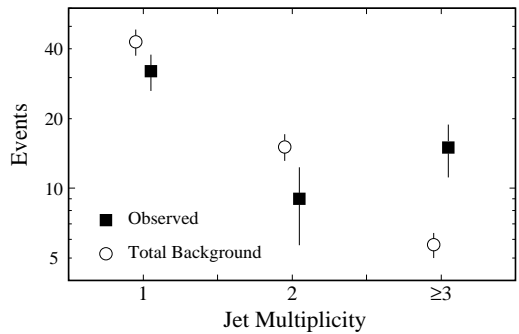,width=70.0mm}}}
\vskip -0.5cm
\caption{Jet multiplicity distribution for single lepton $b$-tag events
(before applying $ \cal {A} $ and $H_T$ cuts), compared to background estimates.}
\vskip -0.3cm
\label{fig:tag}
\end{figure}

The offline event selection in the $b$ tagging analysis
requires one isolated lepton with $E_T > 20 \ {\rm GeV}$
with $|\eta_e| < 2.0$ or $|\eta_\mu| < 1.7$, \met $ > 20 \ {\rm GeV}$,
and 3 or more jets with $E_T > 20 \ {\rm GeV}$ with
$|\eta_{jet}| < 2.0$. Since we require a tag muon in the event, we use
looser cuts on aplanarity and $H_T$
compared to the event shape analysis:
$ \cal {A} > $ 0.040 and $H_T > 110 \ {\rm GeV}$.
Figure~\ref{fig:tag} shows the distribution of the jet multiplicity for background
events and for single lepton data before the $ \cal {A}$ and $H_T$ cuts.
The data agrees with the number of events from $W$ + jets and
QCD processes in the low jet multiplicity region.
For the high multiplicity bin we can see a clear excess above background
even without the $ \cal {A}$ and $H_T$ cuts.

The main background in the $b$ tagging analysis is also $W$+jets production.
The background is estimated by applying the muon tag rate, which is determined by the
fraction of jets tagged in multijet events, to the jets in the background
sample after all other cuts described above are applied.

Eleven events passed the above selection criteria in the $b$ tagging analysis
of the single lepton channel, with estimated background $2.4 \pm 0.5$
events (see Table~\ref{tab:all}).

\subsection{ $e\nu$ Decay Channel}

In order to identify \ttbar events in which neutrinos from both $W$
 decays carry much of the transverse momentum, we perform
an $e\nu$ decay channel analysis.
This analysis is mainly focused on selecting those \ttbar events
that fail the event selection criteria
for the $ee$, $e\mu$ and $e$+jets decay channels.
The signature of the \ttbar events in the $e\nu$ decay channel is
one high $p_T$ isolated electron, two or more jets and
very large \met which together with the electron, forms a transverse mass
much higher than the $W$ mass.

The event selection requires
events with an isolated electron with
$E_T > 20 \ {\rm GeV}$ with $|\eta_e| < 1.1$ and at least two jets with
$E_T > 30 \ {\rm GeV}$ with $|\eta_{jet}| < 2.0$. We require that the
\met is above $50 \ {\rm GeV}$ and the transverse mass of the \met and the
electron is above $ 115 \ {\rm GeV}$.
The dominant background in this
channel is $W$+jets production, $W$ pair production and
a misidentified electron with \met in QCD multijet events.

Four events survived the selection criteria for the $e\nu$ decay channel,
with estimated bachground $1.2 \pm 0.4$
events (see Table~\ref{tab:all}).

\subsection{ All-jets Decay Channel}

    The decay of the top quark through the multi-jets channel represents the dominant decay mode in the
standard model (44\%).  But, without the signature of an isolated lepton in the event, the
QCD multi-jets background overwhelms the signal by roughly 1:2000 before any other cuts are applied.
This has made the search for top in this channel very challenging.
Significant gains in this channel have been made by employing multivariate techniques
and a number of novel techniques which are optimized for this search.
The current analysis~\cite{d0alljets} includes the full Run 1 data set.

Artificial neural networks constitute a powerful nonlinear extension of conventional methods of
multi-dimensional data analysis \cite{neural}.  They have been proven extremely well suited to the
multi-jets search because they handle information from a large number of inputs and
properly account for the nonlinear correlations between these inputs.
We note that the output of the neural net is simply a mapping between the multi-dimensional space of
variables describing our events and a one-dimensional output space.  The analysis may proceed by
setting a threshold on the neural net output, corresponding to a cut surface in multi-dimensional space.

This analysis relies on a set of simple topological variables
which discriminate between the fundamental properties of the $t\bar{t}$ signal and the QCD background.
For the background, these properties are:

\begin{itemize}
\item The events have a lower overall energy scale; leading jets have lower
$E_T$; multi-jet invariant masses are small.
\item The additional radiated jets are soft (have low $E_T$).
\item The event topology is more planar (less spherical than top).
\item The jets are more forward-backward in rapidity (less central than top).
\end{itemize}

We employ 10 kinematic parameters which embody these properties.
In addition to the kinematic parameters, we use three other parameters
based on jet widths, $p_T$ of the tagging muon and the mass likelihood.
This set of 13 variables was used as the set of inputs to the neural net.
We used the back-propagation learning algorithm in Jetnet \cite{jetnet} to train the network
and provide a single variable NN (the neural network output in the range (0,1)).

At the simplest level, each $t(\bar{t})$ decays into a $b(\bar{b})$ and $W^{+}(W^{-})$ boson, and the $W$
bosons into light quarks.  Barring extra gluon bremsstrahlung, this represents six quark-jets in
the final state.  The average jet multiplicity for {\sc Herwig} $t\bar{t}$ events ($m_t$=180 GeV/c$^2$)
is 6.9, implying that the contribution from gluons is relatively small.
Conversely, the QCD multi-jet background has final-state jets that originate
predominantly from underlying gluon radiation.  Although gluon splitting can take place,
producing both quark and gluon jets, it is expected that there will be a large
enhancement in gluon content for QCD jet production.  A Fisher discriminant, based on the difference
between quark and gluon jet widths is used to differentiate between $t\bar{t}$ signal and QCD background.

The $p_T$ of the tagging muon helps to further differentiate between signal and QCD background.
Not only does the fragmentation of $b$ quarks produce higher $p_T$ objects, but the $b$ itself is
more boosted in top events than in background.  Thus, the muon $p_T$ spectrum is significantly
harder in $t\bar{t}$ events.  The muon $p_T$ serves as a useful tool in differentiating between
signal and background.

A mass log-likelihood variable provides discrimination between signal and background by
requiring that there are combinations of jets consistent with being from two top decays
and the subsequent $W \rightarrow qq$.
The presence of two $W$'s in $t\bar{t}$ events provides significant rejection against QCD background.
A further requirement that the two reconstructed top quarks have equivalent masses provides
additional discrimination.


\begin{figure}
\centerline{\psfig{figure=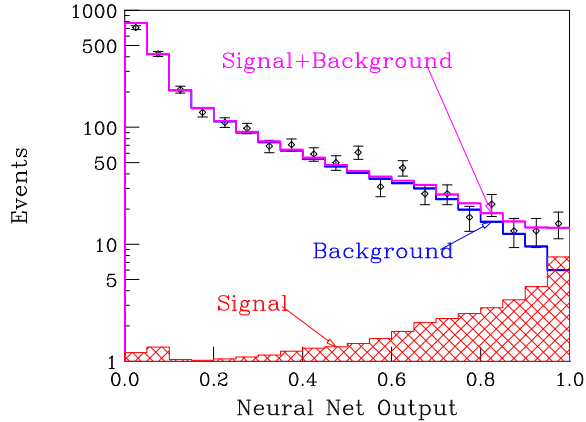,width=3in}}
\caption[Combined fits]
{\small Fit to the distributions in neural net output for signal and background.  The normalizations of
each are fitted to the data (diamonds).}
\vskip -1.3cm
\label{comb_fit}
\end{figure}

The background NN distribution is based upon data events, selected by identical criteria as the signal sample 
except that a tagging muon is not required. To these events, a muon is added with probability and kinematic distributions
as observed in our data. The \ttbar NN distributions are taken from the {\sc Herwig} MC, and checked with {\sc Isajet}.

Once the final neural net has been trained, the relative amount of signal and background in the data may be determined
by a fit for the amounts for signal and background.  Because the shapes of the expected signal and background
distributions in neural net output differ significantly, the relative amounts of each may be disentangled.  This serves
as a useful check of the overall tag rate function and simultaneously provides the top cross section.
This technique benefits from the fact that additional information on the top cross section is extracted even
in regions where the background level is large.  The data (diamonds), and the expected distribution for background and for
signal are shown in Fig.~\ref{comb_fit} along with the combined fit.
The result is, $\sigma_{t\bar{t}}$ = 7.1~$\pm$ 2.8~$\pm$ 1.5 pb at m$_t$ = 172.1 GeV/c$^2$,
in good agreement with the D\O\ result for the leptonic channels~\cite{d0xsec}.
Applying a cut on the neural net output of 0.85 (events shown in Table~\ref{tab:all}), and using the absolute
background predicted, yields a consistent cross section.
Additionally, a more restrictive cut on the neural net output of 0.94, chosen for maximum expected significance using
Monte Carlo data, leaves 18 events with an expected background of 6.9~$\pm$ 0.9.  The probability of an upward
fluctuation of the background to the observed signal is equivalent to 3.2 standard deviations.

\subsection{ \ttbar Production Cross Section}

\begin{table*}[thb]
\setlength{\tabcolsep}{1.5pc}
\newlength{\digitwidth} \settowidth{\digitwidth}{\rm 0}
\catcode`?=\active \def?{\kern\digitwidth}
\caption{Event yields}
\label{tab:all}
\begin{tabular}{lccccc}
\hline
channel           & events          &  estimated  &
\multicolumn{3}{c}{expected signal}     \\
                  & observed        &  background  &  \multicolumn{1}{c}{$m_t$=170 {\rm GeV/c}$^2$}  \\
\hline
$ee$              &  1              & 0.5$\pm$0.1   &  1.2$\pm$0.2 \\
$e\mu$            &  3              & 0.2$\pm$0.2   &  2.2$\pm$0.5 \\
$\mu\mu$          &  1              & 0.7$\pm$0.3   &  0.6$\pm$0.1 \\
$e$+jets(event shape) &  9          & 4.5$\pm$0.9   &  8.6$\pm$1.6 \\
$\mu$+jets(event shape) &  10       & 4.2$\pm$1.0   &  5.5$\pm$1.7 \\
$e$+jets($b$-tag) &  5              & 1.1$\pm$0.4   &  3.6$\pm$0.6 \\
$\mu$+jets($b$-tag) &  6            & 1.4$\pm$0.2   &  2.3$\pm$0.5 \\
$e\nu$            &  4              & 1.2$\pm$0.4   &  1.7$\pm$0.5 \\
\hline
total (leptonic channels) &  39             &13.7$\pm$2.2   & 25.7$\pm$4.6 \\
\hline
all-jets          &  41             &24.8$\pm$2.4   & 15.9$\pm$2.6 \\
\end{tabular}
\end{table*}

Table~\ref{tab:all} shows the number of observed events for all
nine decay channels,
the estimated number of background events, and
the expected number of \ttbar events for $m_{top} = 170 \ {\rm GeV/c}^2$
(\ttbar production cross sections using
ref.~\cite{theory}).

The top production cross section in the leptonic channels 
is calculated using the formula
$
\sigma_{t\bar t} = {{\sum_{i=1}^8 { \left( N_i - B_i \right)}}
 \over {\sum_{i=1}^8 {\varepsilon_i {\cal B}_i L_i}}}
$
where $N_i$ is the number of observed events, $B_i$ is the
expected background, $\varepsilon_i$ is the detection
efficiency for top, ${\cal B}_i$ is the branching ratio and $L_i$ is
integrated luminosity for channel $i$, where $i$=1,...,8.
This cross section is then combined with the aforementioned cross section
measured in the all-jets decay channel taking into account appropriate correlations.

\begin{figure}[hbt]
\vskip 0.1cm
\centerline{\hbox{
\psfig{figure=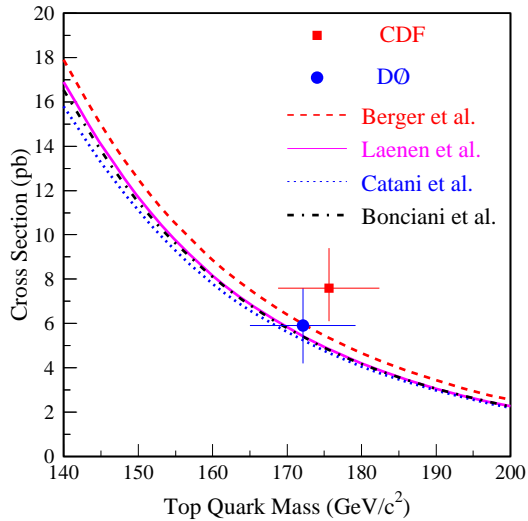,width=70.0mm}}}
\vskip -0.2cm
\caption{ Top cross section vs top quark mass }
\vskip -0.2cm
\label{fig:xs}
\end{figure}

Figure~\ref{fig:xs} shows the D\O\ \ttbar cross section as a
function of top
quark mass compared with theoretical predictions.
We measure the \ttbar production cross section
to be $\sigma_{t\overline{t}} = 5.9\pm1.7$ pb at
our measured top quark mass of $m_{top}=172.1$ GeV/c$^2$
(see section~\ref{sec:mass})
in good agreement with the standard model predictions~\cite{theory,Berger,Catani,Bonciani}.

\begin{figure}[hbt]
\centerline{\hbox{
\psfig{figure=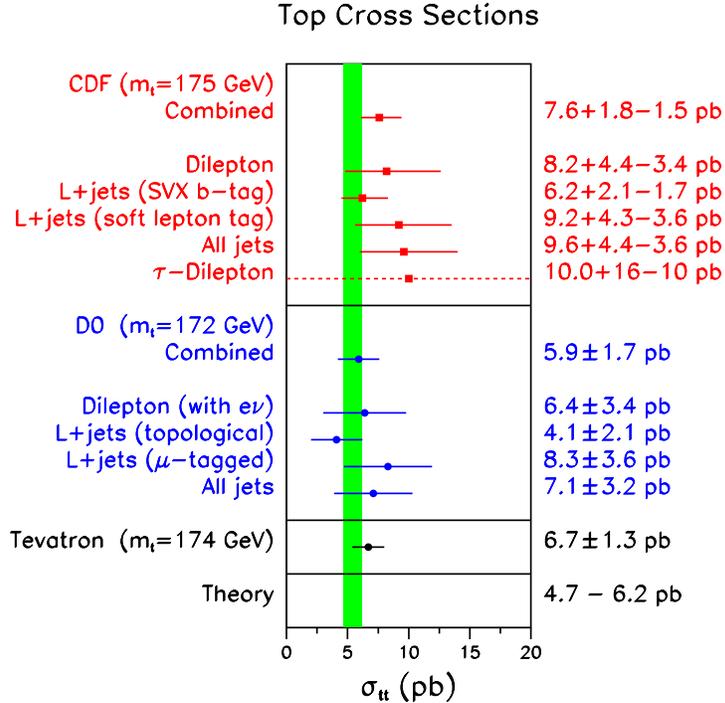,width=100.0mm}}}
\vskip -0.3cm
\caption{ Summary of \ttbar cross section measurements and theoretical calculations }
\vskip -0.2cm
\label{fig:xs_sum}
\end{figure}

Combining the D\O\ and the CDF~\cite{velev}  \ttbar measured cross sections taking into account
correlations between systematic errors yields unofficially
$\sigma_{t\overline{t}} = 6.7\pm1.3$ pb at
an averaged top quark mass of $m_{top}=174$ GeV/c$^2$. Figure~\ref{fig:xs_sum}
shows the D\O\ and the CDF measured \ttbar cross sections as well as the unofficial
Tevatron average and a theoretical band which contains the four calculations.

\section{ Measurement of the Top Quark Mass}
\label{sec:mass}
Direct measurements of the top quark mass using single lepton and dilepton
events are reported here.

\subsection {Single Lepton Events}

\subsubsection {Event Selection}

The initial event selection for the top quark mass measurement analysis
is similar to that used in the cross
section analysis for the single lepton channel, but without
$ \cal {A}$ and $H_T$ cuts.
We require one isolated lepton, $e$ or $\mu$, with
$E_T > 20 \ $ GeV and $|\eta_e|<2.0$ or $|\eta_{\mu}|<1.7$. We also require
\met $> 25 $  {\rm GeV} for $e$ + jets and 20 {\rm GeV} for $\mu$ + jets.
Only the events with four or more jets having $E_T > 15 \ $ GeV with
 $|\eta_{\rm jet}| < 2.0$ are used.
To suppress background in untagged events, we require $E_T^L > 60$
GeV and $|\eta_W| < 2$ for the $W \rightarrow
\ell \nu$.

Ninety events passed the event selection requirements. Among these,
seven events are $b$-tagged events.

\subsubsection {Fitting Algorithm}
For each event passing the above selection cuts, we make a two constraint
kinematic fit~\cite{sss} to the
\ttbar $ \rightarrow \ell + {\rm jets}$ hypothesis by minimizing a
$\chi^2=(v-v^*)^TG(v-v^*)$, where $v(v^*)$ is the vector of the
measured (fit) variables and $G^{-1}$ is its error
matrix~\cite{d0masslj}. Both reconstructed $W$ masses are constrained
to equal the $W$ mass and we assume both
$t$ and $\overline{t}$ quarks have
the same fit mass, $m_{\rm fit}$.
Kinematic fits were performed on all permutations of the jet assignments
of the four highest $E_T$ jets,
with the provision that muon-tagged jets were always assigned to a $b$-quark
in the fit.
Each fit yields a fitted mass value, $m_{\rm fit}$ and a $\chi^2$. The fit with the
lowest $\chi^2$ is chosen to describe the event. Only the events with
$\chi^2 < 10$ are used in the top quark mass determination. 77
events passed the $\chi^2$ cut and among them five events are $b$-tagged
events.
Although $m_{\rm fit}$ is strongly correlated with the top quark mass,
$m_{top}$,
$m_{\rm fit}$ is not the same as $m_{top}$, because of
gluon radiation and permutation ambiguities.

\subsubsection {Mass and Error Determination}
\begin{figure}[hbt]
\centerline{\hbox{
\psfig{figure=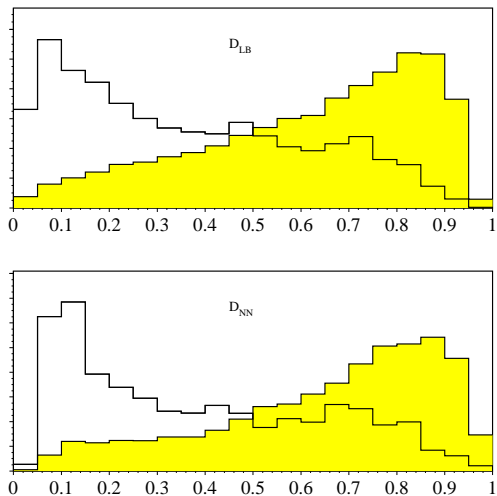,width=70.0mm}}}
\vskip -0.5cm
\caption{ Distribution of discriminants, $D_{LB}$ and $D_{NN}$
for 175 GeV/c$^2$ top MC (shaded histogram) and for background
(unshaded histogram).}
\vskip -0.2cm
\label{fig:discr}
\end{figure}

To further separate the signal and background events we use variables that
can provide good separation between \ttbar and background events
without much correlation to the fitted mass. The following four
variables are
chosen to compute the top quark likelihood discriminant ($D$):
\begin{itemize}
\item \met.
\item $ \cal {A}$.
\item $H_{T2} / \Sigma |p_Z|$, where $H_{T2}$ is the $H_T$ without the $E_T$
of the leading jet.
\item min$(\Delta R_{jj}) E_T^{min} / (E_T^L)$,
where $(\Delta R_{jj})$ is the minimum $\Delta R$ between all
pairs of the jets and $E_T^{min}$ is the smaller jet $E_T$ from the minimum
$\Delta R$ pair.
\end{itemize}
D\O\ uses two discriminants~\cite{d0masslj}.
One is the
$D_{LB}$ (low bias) method, in which we parametrize
${\cal L}_i(x_i) \equiv s_i(x_i)/b_i(x_i)$, where $s_i$ and $b_i$ are the top
signal and background densities in each variable, integrating over the others.
We then form the log likelihood $\ln{\cal L} \equiv \sum_i \omega_i
\ln{{\cal L}_i}$, where the weights $\omega_i$ are adjusted slightly away
from unity to nullify the average correlation (``bias'') of ${\cal L}$ with
$m_{\rm fit}$. For each event we set $D_{\rm LB} = {\cal L} / (1 +
{\cal L})$.
The data are then divided into two bins:
a low signal-to-noise bin and
a high signal-to-noise bin, according to
whether the LB cut is passed.
The LB cut is passed if either $D_{LB} > 0.43$ and
$H_{T2} > 90$ GeV, or if
a $b$ tag exists in the event.  Another method uses a
neural network~\cite{nn} with
the same four variables as input, five hidden nodes, and one node with output
$D_{NN}$.  We divide the data into ten bins in $D_{NN}$.
Figure~\ref{fig:discr} shows the distribution of the discriminants
$D_{\rm LB}$ and $D_{\rm NN}$ for signal and background. They indicate that
either discriminant provides good discrimination.

Since the selected event sample contains both
\ttbar and background events,
we make a two dimensional likelihood fit of the event sample
to the sum of the expected \ttbar signal plus background in the
$m_{fit}$ vs. $D$ plane.
We make an independent likelihood fit for each top quark mass
hypothesis. We use the {\sc herwig}~\cite{herwig} MC to
simulate \ttbar events.
We estimate background using
a combination of $W$+jets events from the {\sc vecbos}~\cite{vecbos}
MC and fake lepton events obtained directly from D\O\ data.

\begin{figure}[thb]
\centerline{\hbox{
\psfig{figure=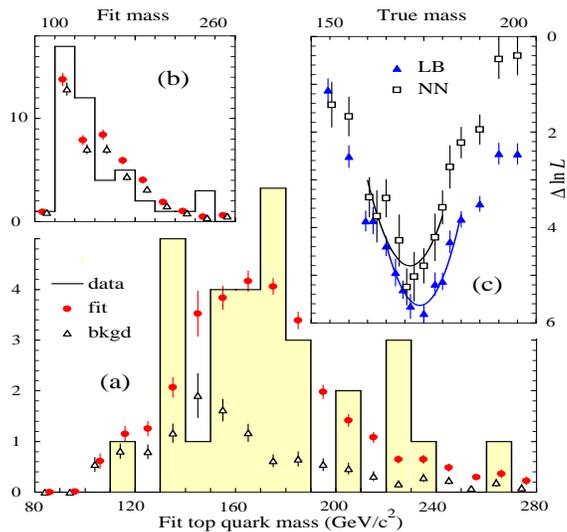,width=75.0mm,height=70.0mm}}}
\vskip -0.2cm
\caption{ (a-b) Fit top quark mass distribution for
events (a) passing or (b) failing the LB cut.
(c) Log of likelihood $L$ {\it vs.}
true top quark mass $m_{top}$ for both LB and NN fits.
}
\vskip -0.1cm
\label{fig:topmass}
\end{figure}

Figures~\ref{fig:topmass} (a) and (b) show the distributions of $m_{\rm fit}$ for
data (a) passing and (b) failing the LB cut. The histograms are data, the
dots are the predicted mixture of signal plus background, and triangles
are background.
Figure~\ref{fig:topmass} (c) shows the log of the
fit likelihood $L$ {\it vs.}
true top quark mass $m_{top}$ for both the LB and NN fits.
The curves are quadratic fits to the lowest point and its 8
nearest neighbor points. The minimum position of each curve
yields the measured top quark mass. The width of the curve
at 0.5 above the minimum determines the statistical error of the
measurement.
The LB fit yields $m_{top} = 174.0 \pm 5.6 \; \rm{(stat.)} $ GeV/c$^2$.
The NN fit yields $m_{top} = 171.3 \pm 6.0 \; \rm{(stat.)} $ GeV/c$^2$.

There are several
sources of systematic error~\cite{d0masslj,d0massPRD}. The major
uncertainties are from the jet energy scale and the MC modelling.
We assign a jet energy scale error of $\pm$(2.5\% + 0.5 GeV)
based on a detailed study of $\gamma$+jet events in data
and MC, particularly focused on the dependence of the $E_T$ balance upon
$\eta$ of the jet, and checked by
the $E_T$ balance in $Z$+jet events.
This leads to an error on $m_{top}$ of
$\pm$4.0 GeV/c$^2$.
The uncertainties in the MC modeling of top and $W$+jet production
are estimated to have a systematic error on the top mass of
$\pm$3.1 GeV/c$^2$.  Other effects,
including calorimeter noise,
multiple $p{\bar p}$ interactions, and differences in fits to $\ln{L}$,
contribute $\pm$2.2 GeV/c$^2$.   All systematic errors sum
in quadrature to $\pm$5.5 GeV/c$^2$.

\vskip -0.1cm

Combining $m_{top}$ from both methods, LB and NN,
we determine the top quark mass to be
$m_{top} = 173.3 \pm 5.6 \; \rm{(stat)} \pm 5.5 \; \rm{(syst)}$ GeV/c$^2$
or $m_{top} = 173.3 \pm 7.8$ GeV/c$^2$,
allowing for the ($88 \pm 4$)\% correlation between two methods.

\subsection{Dilepton Events}

\begin{figure}
\begin{minipage}[b]{.46\linewidth}
 \centering\psfig{figure=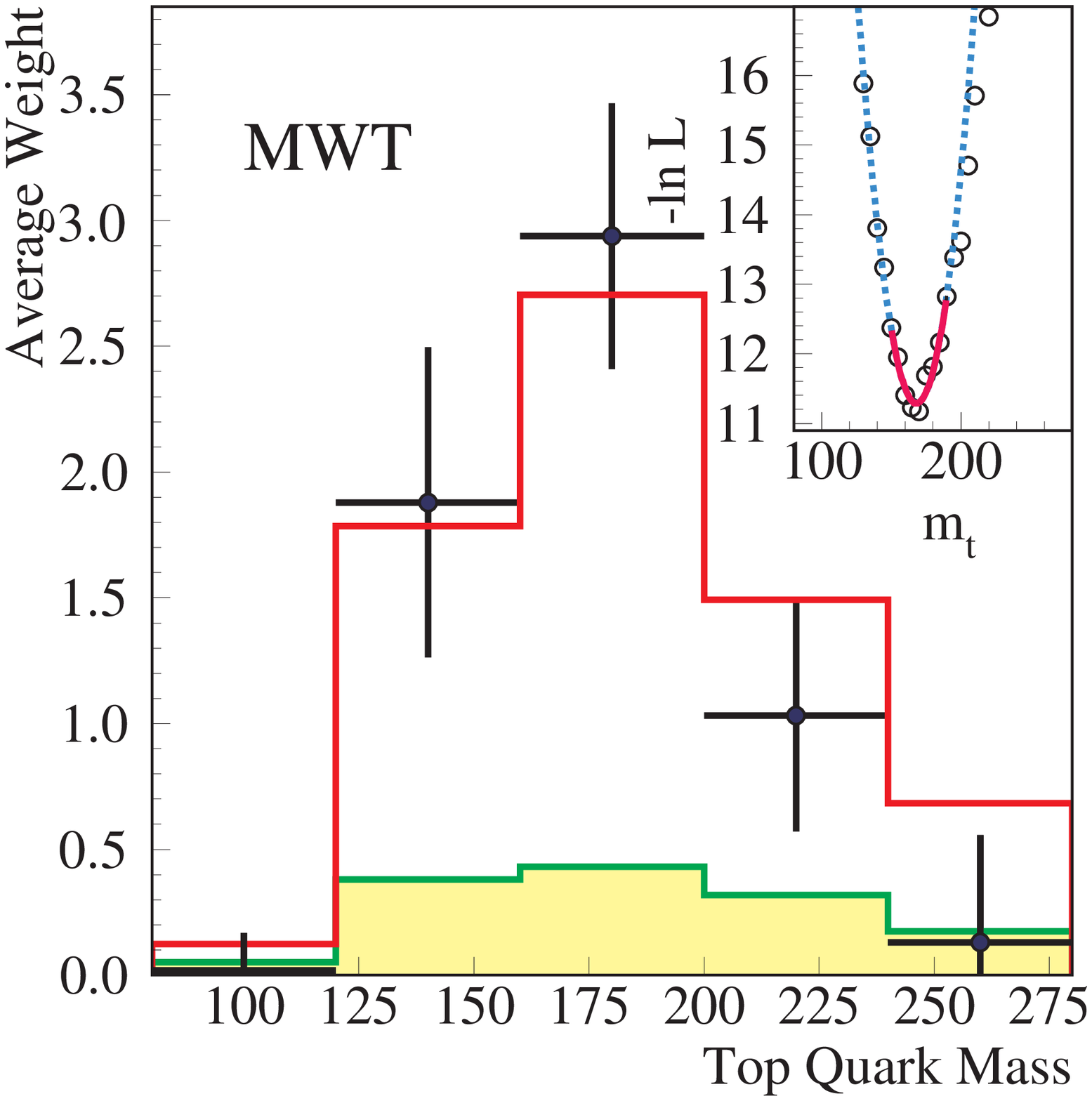,width=2.5in}
\end{minipage}\hfill
\begin{minipage}[b]{.46\linewidth}
 \centering\psfig{figure=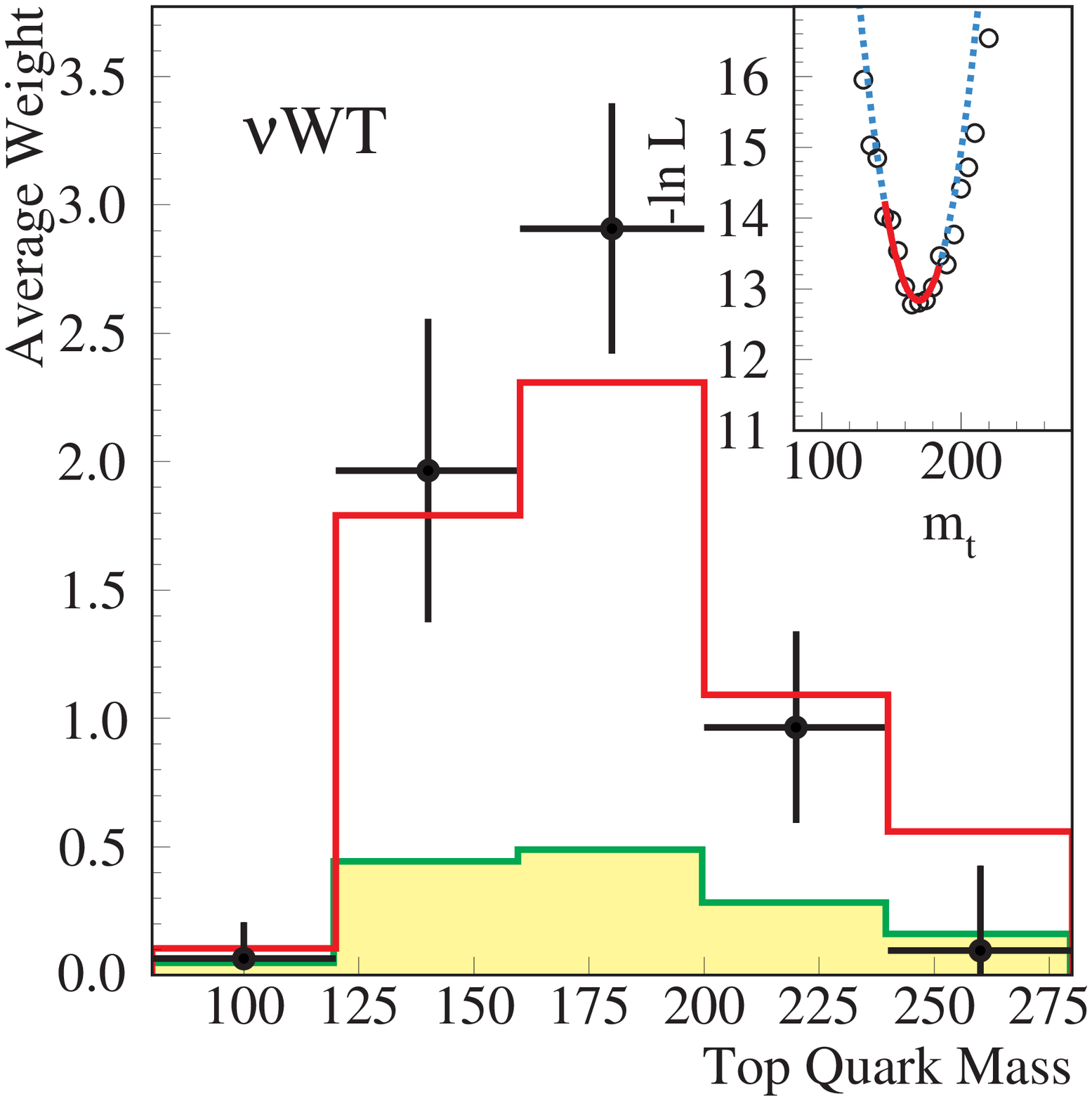,width=2.5in}
\end{minipage}
\caption{Sum of the normalized candidate weights grouped into the five bins
considered in the maximum likelihood fit (circles) for the D\O \ MWT (left)
and $\nu$WT (right)
analyses. The uncertainty on these points is taken from the RMS spread of the
weights in MC studies. Also shown are the average weights from the best-fit
background (dashed) and signal plus background (solid). The -ln L distributions
and quadratic fits are inset.}
\vskip -0.9cm
\label{fig:msd0ll}
\end{figure}

Due to the presence of two neutrinos, dilepton events do not contain
enough information for a constrained fit.
Therefore, we resort to using mass estimators other
than the reconstructed mass. The same
technique that was used for the lepton plus jets mass analyses
is used to calculate a top quark mass likelihood in the dilepton mass
analysis.
D\O\  supplies the missing constraint in the problem by assuming a top
quark mass and reconstructing the event for every assumed
mass~\cite{d0massll}. Based on the reconstructed final
state a weight is computed which characterizes the probablity for this 
event to be from
a $\ttbar$ decay with the assumed mass. There are two algorithms
used to determine the weight. The matrix element weighting (MWT) method
uses the proton structure functions and the probability density function
for the energy of the decay lepton in the rest frame of the top quark
(an extension of Ref.~\cite{KDG}). The neutrino weighting method
($\nu$WT) assigns the weight based on the available phase space for the
neutrinos, consistent with the measured \met.
A maximum likelihood fit is performed to the
shape of the normalized weight curve summed over all 6 dilepton mass events 
using MC derived
probability density functions for signal and background
(see Fig. \ref{fig:msd0ll}). The results for the two analyses are
in excellent agreement :
$m_t = 168.1\pm 12.4 (\hbox{stat}) \ {\rm GeV}/c^2 $ (MWT) and
$m_t = 169.9\pm 14.8 (\hbox{stat}) \ {\rm GeV}/c^2 $ ($\nu$WT)
with a systematic error of 3.7$ \ {\rm GeV}/c^2 $.
By combining the two results, taking into account
the correlations (77\%), we determine the
top quark mass in the dilepton channels to be
$m_t = 168.4\pm 12.3 (\hbox{stat}) \pm 3.6(\hbox{sys}) \ {\rm GeV}/c^2$
or $m_t = 168.4\pm 12.8 \ {\rm GeV}/c^2$.

D\O\ has also combined its dilepton and lepton+jets mass measurements
into a single top quark mass measurement of
$m_t = 172.1\pm 5.2 (\hbox{stat}) \pm 4.9(\hbox{sys}) \ {\rm GeV}/c^2$ 
or $m_t = 172.1\pm 7.1 \ {\rm GeV}/c^2$.

\subsection{Combined Top Quark Mass }

\begin{figure}[hbt]
\centerline{\hbox{
\psfig{figure=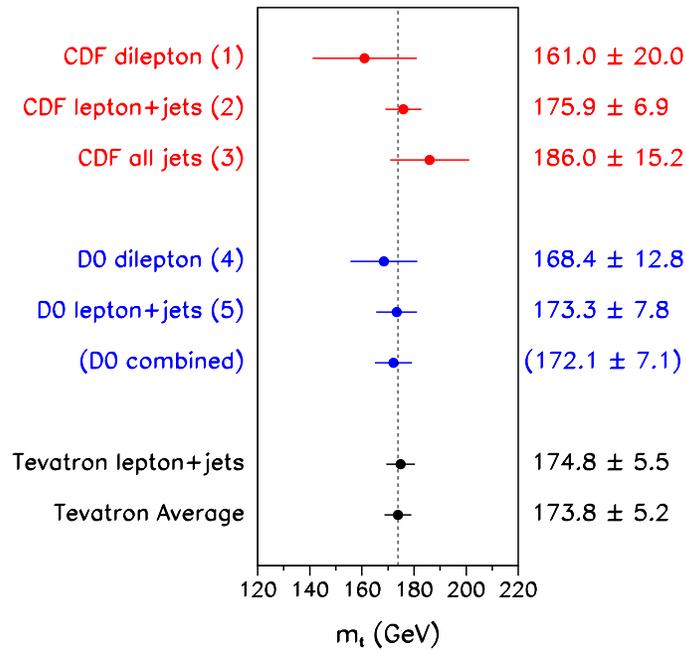,width=90.0mm}}}
\vskip -0.5cm
\caption{ Summary of direct measurements of the top quark mass }
\vskip -0.2cm
\label{fig:mass_sum}
\end{figure}

Combining the D\O\ and the CDF~\cite{velev} top quark mass measurements taking into account
correlations between systematic errors yield unofficially $m_{top} = 173.8 \pm 5.2$ GeV/c$^2$.
Figure~\ref{fig:mass_sum}
shows the D\O\ and the CDF measured top quark mass as well as the unofficial
Tevatron averages.

\section{Charged Higgs Search}

\subsection{Introduction and Strategy}

The ``minimal'' Higgs sector of the Standard Model (SM) is comprised of only
one complex Higgs doublet,
which manifests itself in a single physical neutral Higgs scalar $H^{0}$
($m_{H^{0}}$ is a free parameter of the SM).
However, the scalar sector of the SM is not yet constrained by
experiment, and it is possible that the Higgs content is richer and its
structure more complicated.
A strong motivation for extending the Higgs sector comes from supersymmetry,
where two Higgs doublets are needed to provide masses to the charged fermions.
Two-higgs doublets appear in many theories, both in the context of the SM
and extended models such as the $E_{6}$ superstring-inspired model.
Charged Higgs bosons appear in these as well as in several more general
theories \cite{hhg}.

The minimal extension to the Higgs sector has two complex $SU(2)_{L}$
(isospin $T_3$) Higgs doublets with opposite hypercharge \cite{Drees}.
In the Minimal Supersymmetric Model
(MSSM) there are three parameters:
the mass of the $W$ boson ($m_W$),
the ratio of the two neutral Higgs VEVs
($\tan$ $\beta$),
and the mass of the charged scalar ($m_{H^{\pm}}$).
If kinematically allowed, {\rm BR}(\tHb) can be comparable to,
or even larger than, {\rm BR}(\tWb) depending on \mHp and $\tan$ $\beta$.
At the Tevatron one can therefore search for a light charged Higgs boson 
in the decay of top quarks.

There is a direct lower limit from LEP experiments corresponding
to $m_{H^{\pm}} > 75$ GeV at 95\% CL \cite{LEP2}.
Based on their direct and indirect searches, the CDF experiment has recently
excluded additional regions in the [\mHp,\tbp] plane \cite{CDFch1}.
Also, an independent analysis of CDF data led to similar conclusions for
high \tbp \cite{dpr1}.
Our analysis is based on a somewhat different strategy 
and is sensitive to similar 
but not identical regions of the parameter space.

The value of the \ttbar production cross section and
the mass of the top quark from direct observation at \dzero
\cite{d0xsec,d0masslj} are based on
the assumption of BR($t \rightarrow W^+b$) = 1.
The topic of present interest is based on the violation of this assumption.
We therefore cannot use those measurements in this analysis.
Consequently, in the process of setting limits in the [\mHp, \tbp] plane,
\cstt and \mt serve as input parameters.
Production of \ttb takes place via strong interactions, and
the cross section is not affected by the existence of a charged Higgs boson
lighter than the top quark.
One can therefore use a theoretical prediction based purely on the SM for
\csttp.
There is also no rigorous reason to use the directly measured value of \mtp \
when considering a region where ${\rm BR}(\tHb) \approx 1$, but in the
absence of a compelling reason to start elsewhere, one can choose \mt in the
170 - 175 GeV range as a starting point.
In our analysis we have considered so far charged Higgs predominantly 
decaying into the heaviest quarks or leptons, 
i.e. $H^{+}\to c\bar{s}$ or $H^{+}\to \tau^{+}\nu_\tau$,
(and the charged conjugate decays). The branching ratios for these decays depend
on \tbp. Neglecting other decays into $e^{+}\nu_e, \mu^{+}\nu_\mu$ and
$u\bar{d}$, BR($H\to c\bar{s}$)+BR($H\to \tau\nu_\tau$)=1.

The selection criteria of the \dzero standard measurement of \cstt
were optimized for $t \bar{t} \rightarrow W^{+}bW^{-}\bar{b}$.
In the end, the number of observed events in excess of the estimated
backgrounds is interpreted as arising entirely from
$t \bar{t} \rightarrow W^{+}bW^{-}\bar{b}$,
and \cstt was calculated assuming
${\rm BR}(\tWb) = 1$.
In our indirect search, we replace this last assumption with a more general
one:
${\rm BR}(\tWb) + {\rm BR}(\tHb) = 1$.
We then appply the same selection criteria to our simulated signal and
determine, for given values of \mHp, \mt and \csttp, what values of \tbp \ are
unlikely to produce the number of events observed in our data sample
($n_{obs}$).
With selection criteria tuned for high acceptance to
$t \bar{t} \rightarrow W^{+}bW^{-}\bar{b}$ events, the value of
$n_{obs}-\Mean{n_{bkg}}$ is statistically significant, and in excellent
agreement with the predicted \cstt for ${\rm BR}(\tWb) \approx 1$ (\ie,
${\rm BR}(\tHb) \approx 0$).
Therefore, regions of parameter space that imply a high ${\rm BR}(\tHb)$
can be excluded if the acceptance (using our normal selection criteria) for
$t \bar{t} \rightarrow H^+X$ is small, because it is not possible to explain
the large value of $n_{obs}-\Mean{n_{bkg}}$ in these regions.
This is the strategy of our ``disappearance'' analysis.

For the time being, we limit ourselves to the lepton+jets final states,
which accounts for 30 of the 39 \ttb candidate events on which the current
\dzero standard measurement of \cstt is based.
One very appealing feature of the indirect search is that the selection
criteria are already tuned and the non-top backgrounds, as well as the
systematic uncertainties, have been thoroughly studied and are well understood
through the previous analysis.
For an indirect search, one generates the charged Higgs signal using Monte 
Carlo, and
studies its acceptance for the standard event selection criteria.
Events are generated for one channel at a time for any given \mHp.
Subsequently, a sample for a particular value of \tbp can be obtained by
adding events from different channels in a proportion appropriate for that
[\mHp, \tbp].
A single set of selection criteria is applied to all samples, each of which
correspond to a unique [\mHp, \tbp].

\subsection{Results and Discussion}
\label{sc:bayes}

For a given set of the three
parameters (\mtp, \mHp, \csttp), 
we determine the probability distribution of \tbp. There is no generally applicable
strict upper or lower limit on \tbp. However,
validity of the perturbation theory implies upper bounds on the value of the
charged Higgs--fermion Yukawa couplings.
Requirements on the $H^+tb$ couplings yield
an approximate region of phenomenological interest of
$0.2 \leq \tb \leq 200$~\cite{tblim}. These limits are sensitive to \mHp.


We compute minimum and maximum (one-sided) limits on
\tbp at 95\% confidence level.
Since \tbp \ represents a physical parameter, it must have some discrete value
and not a continuous distribution.
It is therefore appropriate to provide minimum and maximum values for \tbp \
separately.
For a given set of [\mtp,\csttp], the maximum allowed
values of \tbp \ can be plotted as a function of \mHp, and,
similarly, the minimum allowed values can be plotted.
These form the boundaries of the regions in the [\mHp, \tbp] plane
that are excluded at 95\% CL.

\begin{figure}
\centerline{\psfig{figure=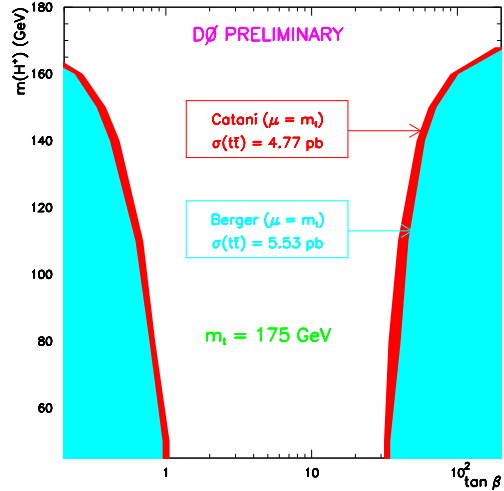,width=75.0mm,height=75.0mm}}
\vskip -0.2cm
\caption{The search results, shown as 95\% CL exclusion contours, in the
[\mHp, \tbp] plane for \mt = 175 GeV and two \ttbar cross section 
calculations.}
\vskip -1.3cm
\label{fig:excl1}
\end{figure}

The lightly shaded region in Fig.~\ref{fig:excl1} represents the region
of [\mHp, \tbp] excluded for \mt = 175 GeV and $\sigma(t \bar{t}) = 5.5$ pb.
The cross section of 5.5 pb is the one calculated by Berger {\it et al.}, for
\mt = 175 GeV, using $\mu = m_{t}$ for the renormalization scale \cite{Berger}.
The reason for choosing this particular cross section is that it is the
largest among the theoretical predictions at this \mtp, and therefore,
for a given number of observed events, the most conservative in terms of
setting exclusion limits on the existence of a charged Higgs.
The darkly shaded region demarcates the additional excluded regions if 
one chooses the cross section calculated by Catani \etal,
($\sigma(t \bar{t}) = 4.8$ pb), for the same renormalization scale
\cite{Catani}.

For \mt = 170 GeV we have
$\sigma(t \bar{t}) = 6.5$ pb (Berger \etal, $\mu = m_{t}$).
According to this cross section, given our lepton+jets signal and
background acceptances, we are most likely to have seen 33 or 34 events if
${\rm BR}(t \bar{t} \rightarrow W^{+}bW^{-}\bar{b}) = 1$.
This means that $n_{obs} = 30$ favors a non-zero contribution from
channels in which at least one of the pair-produced tops decays into a charged
Higgs.
Since the acceptance of the selection criteria for such channels is smaller
than that for $t \bar t \rightarrow W^+ b W^- \bar b$,
the larger the assumed \ttbar production cross section,
the greater the need for a contribution from
a charged Higgs to explain the observed deficit.
A cross section smaller than that found in the standard analysis
would yield a likelihood distribution with a
single peak at $\tb = \sqrt{{m_{t}}\over{m_{b}}} \approx 6$, where
${\rm BR}(t \bar{t} \rightarrow W^{+}bW^{-}\bar{b})$ has a maximum.
As the assumed $t \bar t$ production cross section is reduced, the likelihood
distribution becomes narrower (and the area under the curve decreases), which
results in a larger region of excluded on either side of the peak value.
Like $\sigma(t \bar{t})$, the value of \mt also affects the derived range
of excluded [\mHp,\tbp].

It is important to note that these results are still preliminary. Two aspects of the analysis
are being worked on and will probably change the results. The first one is taking into account
another $H^+$ decay channel, $H^+ \rightarrow W^+b \bar{b}$, which becomes important at 
\tbp $< 2$ once
one considers \mHp $> 130$ GeV or so. Incorporating this additional mode
is expected to somewhat weaken the lower limit on \tbp at large values of \mHp.
The second issue which needs further studies is
the effect of increased widths of top and $H^+$ near the upper and lower boundaries of the range
of \tbp under consideration.

\section{Future Prospects - Run 2}

The next Tevatron collider run (Run 2) is scheduled to begin in the year 2000
with increased luminosity and energy.  D\O\ will increase
its integrated luminosity compared to that accumulated during Run 1
by about a factor of 20 (to
2 fb$^{-1}$).  Combined with the increased Tevatron c.m. energy (from 1.8 to 2.0
TeV) which will yield a 35\% increase in the \ttbar production
cross section and with detector improvements, the number of observed top quark events
will increase by about a factor of 40.
Details of the expected precision for top related measurements during Run 2,
including some which were not done in Run 1 due to the small statistics, are described
in refs~\cite{TeV_2000,Ken_Moriond}.

\section{Conclusions}

D\O\ has made direct measurements
of the top quark mass from single lepton
and from dilepton events using the
entire data sample from the 1992-1996 running period with
an integrated luminosity of 125 \ipb.
The top quark mass is determined to be
$m_{top} = 172.1 \pm 7.1$ GeV/c$^2$.
D\O\ has also measured the \ttbar production cross section.
Thirty nine events were observed in eight different
leptonic decay channels with an estimated
background of $13.7\pm2.2$ events.
In the all-jets mode 41 events were observed with an estimated
background of 24.8$\pm$2.4 events.
The \ttbar production
cross section is measured to be $5.9\pm1.7$ pb at
$m_{top}=172.1$ GeV/c$^2$.

\section{Acknowledgments}

I would like to express my appreciation to the organizers of this
excellent conference for their hospitality.
We acknowledge the support of the US Department of Energy and
the collaborating institutions and their funding agencies in this
work.

\end{document}